# Transition metal nitride thin films deposited at CMOS compatible temperatures for tunable optoelectronic and plasmonic devices

Ryan Bower[1], Daniel A. L. Loch[2], Andrey Berenov[1], Bin Zou[1], Papken Eh. Hovsepian[2], Arutiun P. Ehiasarian[2], Peter K. Petrov[1]

[1] Thin Film Technology Lab, Imperial College London, Royal School of Mines, Exhibition Road, London, SW7 2AZ, UK
E-mail: p.petrov@imperial.ac.uk

[2] National HIPIMS Technology Centre - UK, Materials and Engineering Research Institute, Sheffield Hallam University, Howard St., Sheffield, S1 1WB, UK

## ABSTRACT

Transition metal nitrides have received significant interest for use within plasmonic and optoelectronic devices. However, deposition temperature remains a significant barrier to the integration of transition metal nitrides as plasmonic materials within CMOS fabrication processes. Binary, ternary and layered transition metal nitride thin films based on titanium and niobium nitride are deposited using High Power Impulse Magnetron Sputtering (HIPIMS). The increased plasma densities achieved in the HIPIMS process allow high quality plasmonic thin films to be deposited at CMOS compatible temperatures of less than 300°C. Thin films are deposited on a range of industrially relevant substrates and display tunable plasma frequencies in the ultraviolet to visible spectral ranges. The thin film quality, combined with the scalability of the deposition process, indicates that HIPIMS deposition of nitride films is an industrially viable technique and can pave the way towards the fabrication of next generation plasmonic and optoelectronic devices.

1. Introduction

Plasmonic materials have a wide range of applications, from energy storage and harvesting to biosensing and memory storage devices.[1–4] However, the archetypal plasmonic materials gold and silver are limited in their applicability, displaying poor thermal stability, limited spectral tunability, and incompatibility with standard CMOS fabrication processes. Consequently, transition metal nitrides (TMNs), including titanium nitride and niobium nitride have been suggested as viable alternative plasmonic materials. In addition to superior mechanical and thermal properties compared to gold and silver,[5] TMNs also offer spectral tunability of the plasmonic response, with optical properties dependent on phase, stoichiometry and oxygen impurities.[6] Broadband spectral tunability is observed in ternary transition metal nitrides and is achieved by varying cation ratios.[7] Furthermore, TMNs such as titanium nitride are currently used within CMOS compatible fabrication processes as barrier layers and gate metals.[8–12] However, applications of TMNs are not optimised for optical behaviour and as such, the widespread use of TMNs within plasmonic and optoelectronic devices is limited.

A significant barrier to the application of transition metal nitrides as plasmonic materials within CMOS processes is the prohibitively high temperatures required for optimised thin film optical behaviour (>400°C).[13] This is an area of active research within the plasmonic community and significant strides are being taken to achieve low temperature and CMOS compatible deposition of high-quality TMNs. This has involved the deposition of nitride thin films using a variety of deposition techniques including pulsed laser deposition, DC and RF magnetron sputtering and plasma enhanced atomic layer deposition.[14–16] Deposition by RF sputtering of plasmonic TiN was recently achieved at room temperature and short target-to-substrate distance of 5 cm allowing significant plasma density of the order of $10^{10}$ cm$^{-3}$ and flux of nitrogen ions to reach to substrate.[14]

High Power Impulse Magnetron Sputtering (HIPIMS) can be used to deposit binary and ternary transition metal nitride thin films at CMOS compatible temperatures. HIPIMS utilises high instantaneous powers to produce plasma density of the order of $10^{13}$ cm$^{-3}$ at the substrate which activates the deposition process by increasing the ionisation degree of the sputtered flux and the dissociation rate of the reactive gas ions. Del Giudice et al. have previously deposited micron thick TiNbN films using a hybrid HIPIMS and DC magnetron sputtering method at room temperature.[17] Ehiasarian et al. reported a full morphological densification of TiN thin films[18] due to a high surface mobility and reactivity of the adatom flux and the promotion of (002) crystallographic texture which favours the incorporation of impurities such as oxygen as substitutions or interstitials into the growing crystal lattice of the films and reduces its segregation at grain boundaries.

We present the plasmonic characteristics of nitride films deposited above the threshold for achieving high dissociation and ionisation during deposition which allows access to lower substrate temperatures of 300°C, compatible with CMOS fabrication flow. CMOS compatible transition metal nitride thin films of the form $Ti_{(1-x)}Nb_xN_y$ were deposited on a range of industrially relevant substrates, including steel, MgO, Si, and glass via confocal HIPIMS. Furthermore, we compare the optical properties of co-sputtered ternary transition metal nitrides with multi-layered binary nitrides, as measured by spectroscopic ellipsometry.

2. Results and discussion

**Figure 1** displays the optical properties of $Ti_{(1-x)}Nb_xN_y$ films deposited on steel, Si, glass and MgO substrates at 300°C via HIPIMS, as measured using spectroscopic ellipsometry. Ellipsometric parameters psi and delta were measured at three angles (65°, 70° and 75°) and fit to a Drude-Lorentz model containing two Lorentz oscillators in order to extract the real ($\varepsilon'$) and imaginary ($\varepsilon''$) dielectric permittivity for each film. As can be seen from the real permittivity plots, all films display metallic optical behaviour above 320nm and optical properties comparable to literature values for TiN, NbN and TiNbN.[2,17,19–21]

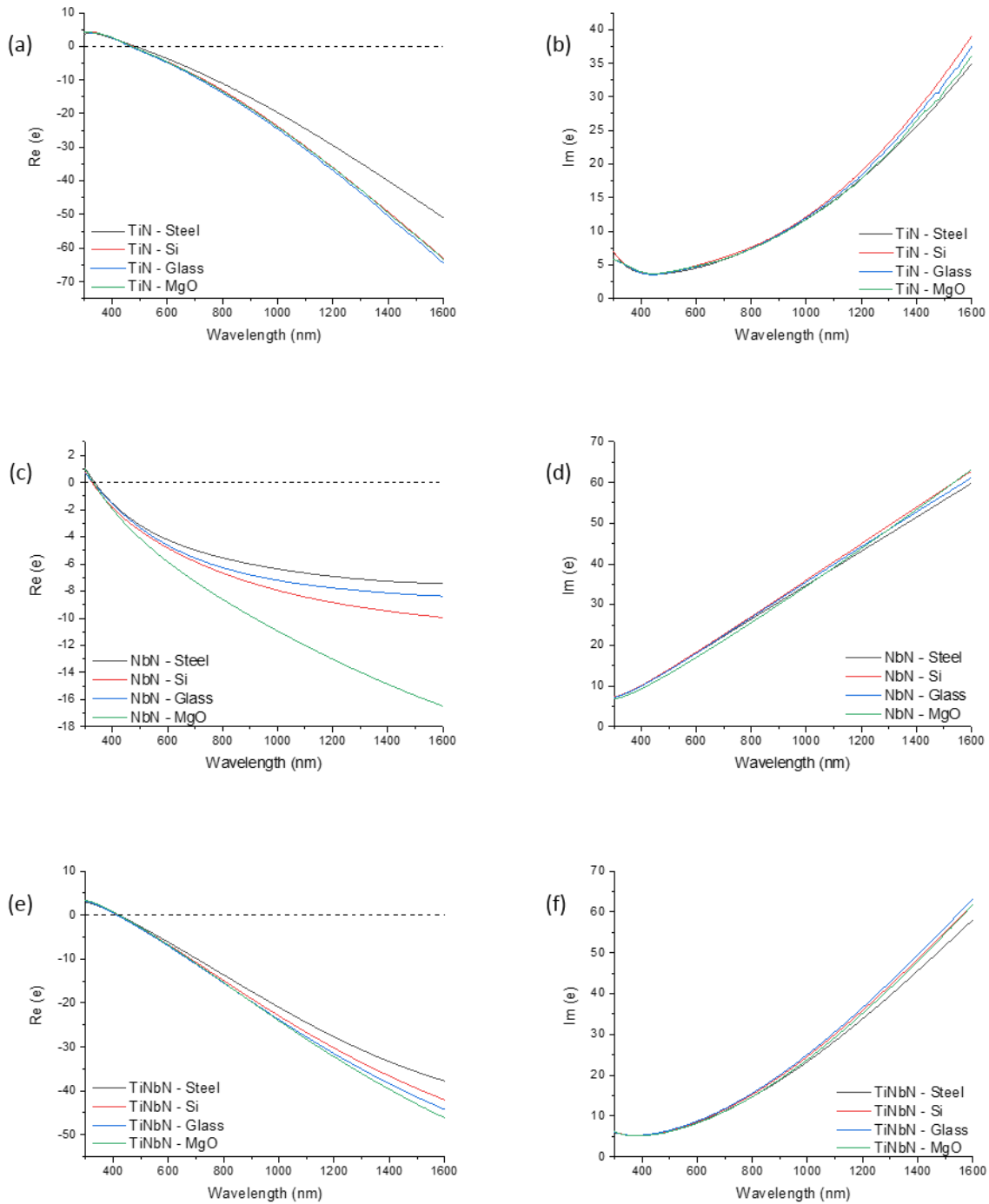

**Figure 1.** Spectroscopic ellipsometry data for TiN (a, b), TiNbN (c, d) and NbN (e, f) thin films deposited on a variety of substrates.

Notably, the thin films were deposited at low, CMOS compatible temperatures (300°C) yet display optical properties comparable to transition metal nitride films deposited at higher temperatures. **Figure 2** indicates that the magnitude of the real dielectric permittivity is comparable to films deposited via RF magnetron sputtering at 600°C on glass and 850°C on sapphire. Films are also of comparable quality to room temperature TiN films deposited with a shorter substrate target distance. Notably, TiN films deposited at 300°C using lower HIPIMS peak powers are shown to have a lower magnitude of ε'.

The binary and ternary films display tuneable optical properties, in addition to highly metallic plasmonic behaviour. This is evident when considering the screened plasma energy, corresponding to the wavelength at which ε' = 0 or the epsilon-near-zero (ENZ) point. For $Ti_{(1-x)}Nb_xN_y$ films this is shown to be tuneable over the range 320 – 490nm. This tunability arises due to variations in charge carrier concentration with cation identity and stoichiometry. The unscreened plasma frequency ($\omega_{pu}$) is related to the conduction electron density ($N_e$) and electron effective mass ($m^*$) as shown in **Equation 1**:

$$\omega_{pu} = \sqrt{\frac{e^2 N_e}{\varepsilon_0 m^*}} \tag{1}$$

Within ternary nitride films, varying the ratio of cations with different numbers of valence electrons allows for the plasma frequency to be tuned, as has previously been reported.[17,22] In addition to producing CMOS compatible TMNs via HIPIMS, we have also deposited layered transition metal nitride structures consisting of alternating layers of binary TiN and NbN thin films. Ellipsometric measurements of these layered films also display negative dielectric permittivities and relatively low losses, falling between the higher losses of NbN and that of TiN, similar to the co-sputtered films, as shown in **Figure 3**. Notably, in contrast to the co-sputtered films, the unscreened plasma frequency of these films is blue-shifted. The quality of the binary, ternary and layered films is directly a result of the HIPIMS deposition technique used. In comparison with the RF sputtering processes that are characterised with a discharge current density of 0.1 Acm$^{-2}$, an average ion energy of ~40 eV and a ratio of dissociated-to-molecular gas ion flux just below 1,[24] the HIPIMS conditions used in the present study were: discharge current density ~0.6 Acm$^{-2}$, yielding a ratio of dissociated-to-molecular gas ion flux above 1, (**Figure S1.2**),[18] whilst retaining a low average ion energy of 3 eV. The HIPIMS process, therefore, carries a lower risk of ion and thermal damage of the substrate. At the same time the higher operating voltage of the HIPIMS process of ~500 V ensures higher sputter yields and faster deposition rates compared to RF sputtering where voltages are <300 V. The plasma composition spectroscopic analysis carried out during the deposition, indicated that both gas dissociation ($N^{1+}/N_2^{1+}$) and metal ionisation ($Ti^{1+}/Ti^0$) ratios exceed 1, (Figure S1.2). The high flux of atomic nitrogen to the surface promotes the formation of low-mobility TiN clusters and effectively traps Ti on (002) oriented grains. At the same time metal ions gain energy through the plasma sheath at the substrate surface and become highly mobile adatoms. Finally, the high powers dissipated in the HIPIMS pulse rarefy the gas atmosphere and reduce energy loss of sputtered atoms on the way to the substrate. These factors work in unison to promote the formation of (002) oriented surfaces. They also lead to the densification of the grain boundaries, smooth the surface and lower the affinity of the coatings towards oxygen.

Increasing the peak power density from 3.8 Wcm$^{-2}$ (DC magnetron sputtering) to 76 Wcm$^{-2}$ (HIPIMS) has been shown to increase the magnitude of the real part of the dielectric function ($\varepsilon'$).[23] Our observations confirm that further extending the range of power densities to 380 Wcm$^{-2}$ enhances plasma density and nitrogen dissociation, (Figure S1.2), and results in significant improvements in the plasmonic properties of all transition metal nitride thin films reported in this study, as shown in **Figure 2 a**.

Increased plasma density is also provided as an explanation for high quality TiN films recently deposited at room temperature via RF magnetron sputtering.[14] These films were deposited with reduced substrate target distance and, as shown in Figure 2, have $\varepsilon'$ comparable to films deposited by HIPIMS. However, the ε'' values for these films are higher than those obtained in this work, which could be attributed to lattice defects due to the high energy of ion bombardment of tens of eV typical of RF excitation.[24] Indeed, when considering the figure of merit for plasmonic materials (-ε'/ε''), we observe that films deposited by HIPIMS are superior to all but epitaxially grown TiN films deposited at high temperatures. We acknowledge that in some cases, lossy plasmonic materials are desired (e.g. for hot electron generation and local heating applications)[2], however, in this instance we are considering the "ideal" plasmonic behaviour of films.

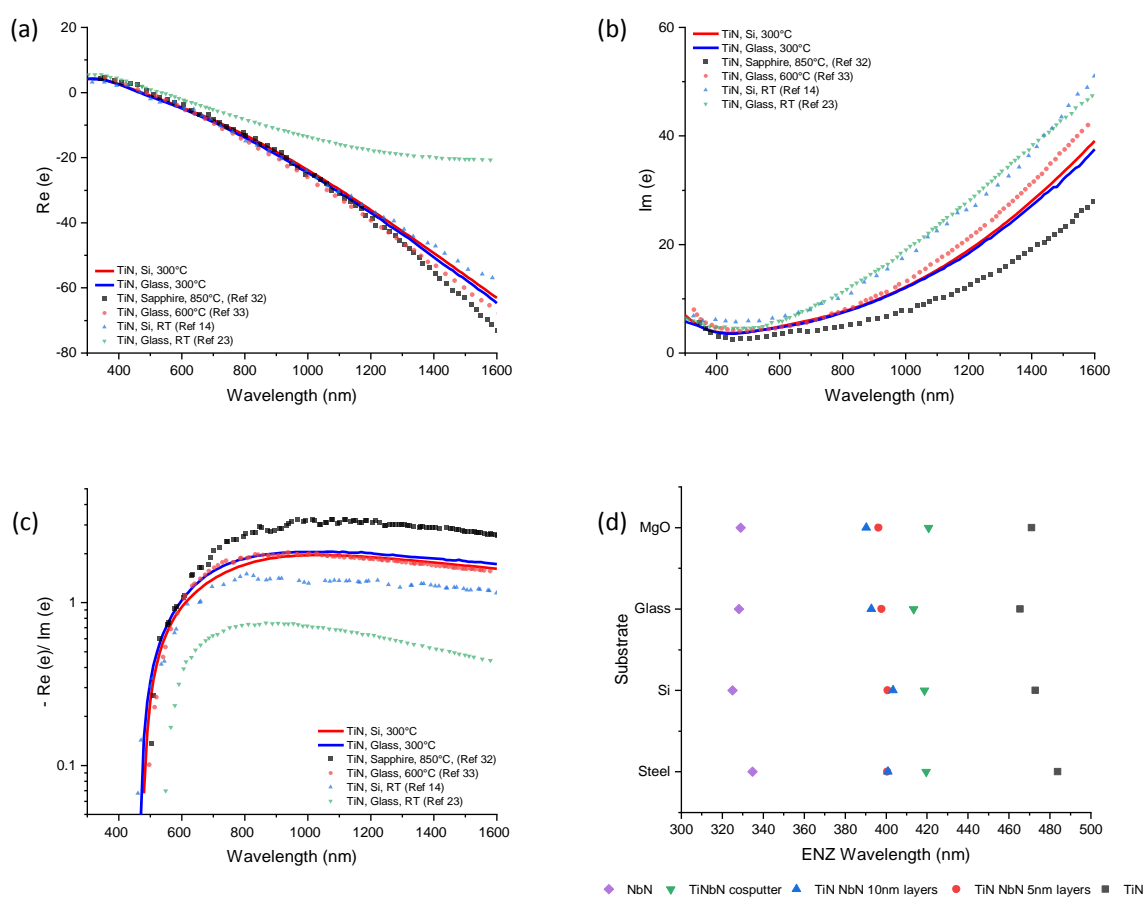

**Figure 2.** Real (a) and imaginary (b) spectroscopic ellipsometry data for TiN films deposited via HIPIMS onto glass and Si substrates. Figures includes a comparison to literature values for TiN films. [14,23,32,33]
(c) Figure of merit (-ε'/ε'') (FOM) comparison of TiN thin films and literature data. References 32, 33 and 14 are deposited via RF magnetron sputtering at 850°C, 600°C and RT respectively. Reference 14 has a short substrate target distance of 5 cm. Reference 23 is deposited by HIPIMS with a peak power density of 75W cm$^{-2}$. All literature films are ~50nm.
(d) A comparison of ENZ crossover wavelength for binary, ternary and layered films.

When considering the optical properties of the binary and ternary thin films, slight substrate dependent variation between films is observed, particularly at longer wavelengths. A number of potential reasons exist for this behaviour, with variations in phase, oxygen content and strain all known to contribute to observed optical and dielectric properties.[25,26]

It has previously been reported for TiN and NbN that phase mixtures and changes in preferred orientation and consequently strain, can result in variations in dielectric permittivity[27,28]. As observed in XRD diffraction patterns,(S3), the thin films deposited are polycrystalline and adopt a cubic rock-salt structure, with dominant preferred orientation along the (111) and (002) directions, in agreement with literature data for TiN, NbN and TiNbN films. Although there is no significant evidence from the XRD diffraction patterns for the presence of alternative phases (e.g. hexagonal NbN or TiN) it is known that certain phases such as hexagonal δ'-NbN have strong peak overlap with cubic δ-NbN at lower diffraction angles. It is therefore not possible to discount the probability that there are small amounts of alternative phases present. These phases could contribute to the slight variation in the dispersion behaviour observed at longer wavelengths. However, this contribution is likely to be minor, as indicated by the Drude – Lorentz behaviour of the binary and ternary thin films.

The presence of small amounts of oxygen impurities and surface roughness of transition metal nitride thin films can also yield distinct differences in the dielectric permittivity measured. As previously mentioned, the low temperature deposition of transition metal nitrides via RF magnetron sputtering has been shown to result in the formation of oxynitride thin films due to residual oxygen partial pressures.[26] In order to minimise this effect, residual gas analysis was performed during deposition and showed low levels of contaminants of OH, O and $O_2$ introduced via the process gas (**Figure S1.1**). Atomic force microscopy measurements of films deposited on Si substrates (S4) indicate that the root mean squared roughness is <1 nm. As such, scattering from surface morphological features is unlikely to contribute to variations in dispersion behaviour. By minimising the effects of oxygen impurities and surface roughness, HIPIMS can produce high quality transition metal nitride thin films.

The tunability of transition metal nitride films is strongly dependent upon film stoichiometry, as shown in the ternary films. These films display unscreened plasma frequencies corresponding to wavelengths ranging from 413-420 nm for all substrates, falling between the average epsilon-near-zero values for NbN and TiN of 330nm and 473 nm respectively. Film composition was assessed via X-ray fluorescence (S2) for films on Si substrates. XRF data indicates cation stoichiometries of x = 0.52 within $Ti_{(1-x)}Nb_xN_y$ films. To further confirm film composition, EDS measurements were performed on co-sputtered $Ti_{1-x}Nb_xN_y$ films on Si substrates. Data were collected at 10 keV using an aperture of 60 microns and indicated that the elements present were niobium, titanium nitrogen and silicon (S2). Neglecting contributions from Si, the stoichiometry of the thin films was calculated to be $Ti_{0.51}Nb_{0.49}N_y$, in good agreement with the XRF data and close to the ratio expected from the deposition conditions.

The ternary films deposited via HIPIMS are of good crystalline quality, containing an alloyed mixture of TiN and NbN, and displaying high quality plasmonic behaviour. The broadband tunability of the ternary TiNbN films indicates the potential for use within optoelectronic devices at a range of functional wavelengths whilst the highly plasmonic behaviour of CMOS compatible films deposited on a variety of substrates suggests the versatility of application that these films can be used for.

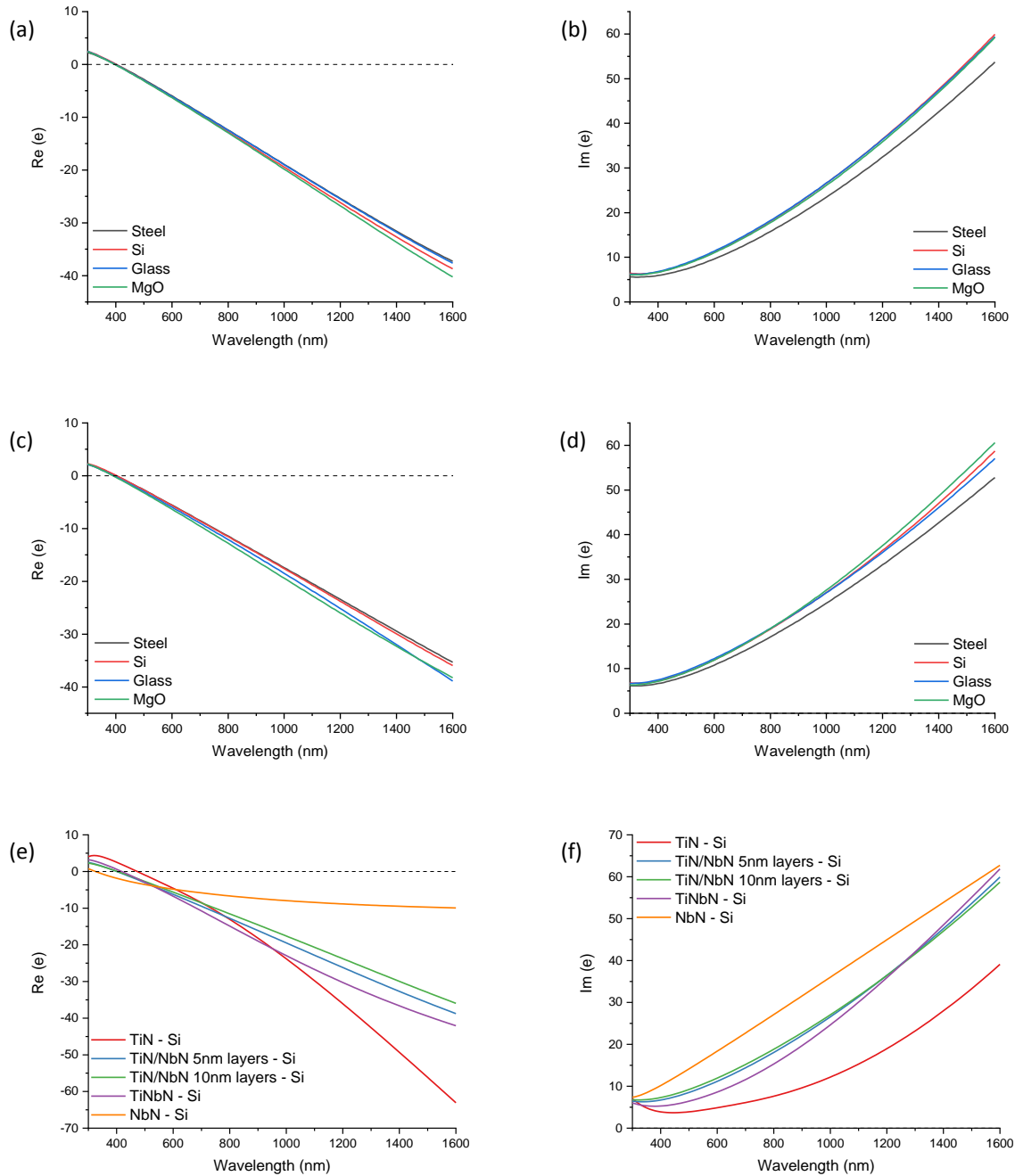

**Figure 3**: Spectroscopic ellipsometry data for (a,b) 5 nm TiN/NbN layers and (c, d) 10 nm TiN/NbN layers.
(e, f) Comparison of ellipsometry data for binary, ternary and layered films on Si substrates. Comparisons of films on other substrates are included in S5.

As shown in **Figure 2 d**, the layered superlattice films display a blue-shifted screened plasma frequency when compared to the co-sputtered films. The layered films have total thicknesses of 100 nm and individual layer thicknesses of 5 nm and 10 nm. As in co-sputtered TiNbN, crystalline film structures with (111) and (002) preferred orientations and surface roughness of less than 1 nm is observed (S4). Whilst some variation in plasma frequency may be expected due to minor variations in film stoichiometry and structure in subsequent deposition runs, the difference in ENZ position is significant, approximately 20 nm. EDX analysis of film stoichiometry, indicates film cation composition of $Ti_{0.52}Nb_{0.48}N$ for both layered films. This is within error range for equivalent cation stoichiometries for each film and therefore does not provide an adequate explanation for the shift in plasma frequency.

A key factor to consider is strain within the layered films. It has previously been reported that ultrathin layers of films of different composition adopt a single, intermediate lattice parameter for each layer.[29] As a result, the films in each layer are strained, in opposite directions. This is in contrast to co-sputtered films which will adopt a single, pseudo crystalline, unstrained unit cell. XRD diffraction patterns collected for the layered films display broad diffraction peaks, however, a single lattice parameter is indeed observed for the layered structures. As mentioned previously, strain within thin films is known to impact the dielectric properties and vary the refractive index.[25] It is our suggestion that the strain within the layered structures accounts for the shift in plasma frequency when compared to co-sputtered films.

Sputtering of ultrathin layers often results in the formation of islands within films unless deposited at high temperatures or epitaxially.[30,31] However, the increased peak power density used within the HIPIMS process allows for continuous films with thicknesses less than 10 nm to be deposited. This is a consequence of adatom mobility due to the high fraction of metal ion flux created by the high power used within the HIPIMS process. This allows layers as thin as 5 nm to be deposited. The high ion-to-neutral ratio and presence of metal ions in the HIPIMS process enhances the mobility and contributes to more homogeneous nucleation, dense grain boundaries and larger grains from the outset of the growth.[18]

This strain mediated tunability offers an additional degree of freedom when considering device design for plasmonics. It enables access to lower plasma frequencies compared to co-sputtered films and at some spectral ranges displays lower losses, an important factor when considering plasmonic performance. Layered structures also offer the option of depositing a metamaterial with a graded effective refractive index by varying the layer thickness throughout the film. Such control is achievable using HIPIMS.

**Conclusion**

We have deposited high quality, plasmonically active binary and ternary transition metal nitride thin films using High Power Impulse Magnetron Sputtering. The deposition process is CMOS compatible and scalable, allowing the manufacture of plasmonic and optoelectronic devices using transition metal nitrides. Deposition is demonstrated to be possible on a range of industrially relevant substrates and films are spectrally tunable within the wavelength range 325-483 nm.

Layered transition metal nitride superlattices are also demonstrated to display high quality plasmonic behavior. Lattice strain within these films results in a blue-shift of plasma frequency, and affords an additional degree of freedom when considering tunability of nitride thin films.

The HIPIMS deposition technique yields high quality plasmonic films due to the high dissociation and ionization achieved during the deposition process. The increased plasma densities yield high quality transition metal nitride films even at low temperatures. This method is suitable for the fabrication of plasmonic devices based on metal nitride thin films. Furthermore, the HIPIMS method of deposition used in the current work is scalable to a target length of up to 3 meters and is preferable for large area samples' production, where uniformity of deposition thickness and film properties are critical.

**EXPERIMENTAL**

Thin films were deposited at the National HIPIMS Technology Centre at Sheffield Hallam University in a CS-400S Cluster system (Von Ardenne Anlagentechnik GmbH) equipped with two confocally-arranged cathodes. Each thin film material was deposited on all substrates types simultaneously. The HIPIMS discharge was operated in a constant current mode using Highpulse 4000 (G2) generators (Trumpf Hüttinger Elektronik sp. z o.o.) at a peak current density of 0.6 A cm$^{-2}$ and peak power density of 380 W cm$^{-2}$. The constant current regulation of the generator allowed for an extremely stable and reproducible discharge with arc rates of ~3 arcs per hour. Residual gas analysis was carried out during the deposition via a differentially-pumped quadrupole mass analyser type Microvision 2 (MKS). Plasma composition analysis was carried out using a plasma-sampling energy-resolved mass spectrometer type PSM3 (Hiden Analytical Ltd.) whose collection was synchronised with the peak of the pulse delivery at the substrate and also in time-averaged mode. Energy scans were integrated to obtain the total ion flux for each species.

Ellipsometry data was collected using a J. A. Woollam W-VASE ellipsometer. Data were collected over the spectral range 300-1600 nm at incident angles of 65°, 70° and 75°. As the films are optically thick, the experimental data was directly fit using the Marquardt minimisation algorithm to a Drude - Lorentz model, using two Lorentz oscillators. Drude - Lorentz parameters are summarised in Table S5.1.

XRD data for each film were collected over the 30°≤2θ≤50° range using a Bruker D2 Phaser X-ray diffraction system operating in the Bragg-Brentano theta-theta geometry with a Cu K$_\alpha$ wavelength of 1.54 Å. Thin film XRD data was collected using a Panalytical Empyrean operating in theta-2theta geometry. The system is equipped with a monochromator and a 2D PIXcel® detector.

Chemical analysis was performed via Energy Dispersive X-ray Spectrometry (EDS) using an Oxford Instruments INCA. Samples were imaged in a Zeiss Leo Gemini 1525 FEG SEM with 10 keV operating voltage and 60 micron aperture. Element quantification was achieved using a titanium standard film, with 97% measurement accuracy. Chemical composition was corroborated using quantitative X-ray fluorescence XRF (Panalytical Epsilon 3 spectrometer) using thin film analysis software (Stratos).

**Acknowledgements**

This work was partly supported by the Engineering and Physical Sciences Research Council (EPSRC) Reactive Plasmonics Programme (EP/M013812/1) and by the Henry Royce Institute through EPSRC grant EP/R00661X/1.

Ryan Bower acknowledges funding from the EPSRC Centre for Doctoral Training in Advanced Characterisation of Materials (Grant Ref: EP/L015277/1)

# SUPPLEMENTARY INFORMATION

1. <u>Deposition parameters and analysis</u>

Films were deposited using the following deposition parameters:

| Film | Number of Operating Cathodes | Peak Current [A] | |
|---|---|---|---|
| | | Ti target | Nb target |
| TiN | 1 | 45 | - |
| TiNbN co-sputter | 2 simultaneously | 45 | 35 |
| TiN/NbN 10nm layers | 1 at a time | 45 | 35 |
| TiN/NbN 20nm layers | 1 at a time | 45 | 35 |
| NbN | 1 | - | 35 |

*Table 1: HIPIMS deposition parameter summary.*

The sputtering target diameter was 100 mm, Ti purity was 99.6% and Nb purity was 99.9%. The target to substrate distance is 90 mm. The substrate temperature was measured using a thermocouple positioned 5 mm behind the substrate holder and calibrated for the surface of a Si substrate. The base pressure was $3\times10^{-6}$ Pa. Nitrogen and argon gas purity was 99.998% (N2.7). For all runs, the total pressure during deposition was maintained at 0.3 Pa by a PID controlled pump throttle valve. The Ar and $N_2$ gas flow ratio $qV(Ar):qV(N_2)$ was kept at 30:1 for single and multilayer depositions. For co-sputtering runs, the gas flow ratio was adjusted to 15:1 to compensate for higher reactive gas consumption of the two cathodes operating simultaneously. The substrates were at floating potential (no bias potential was applied to them). Substrates were chemically cleaned by standard acetone, IPA and deionised water wash in an ultrasonic bath before loading into the cluster system. The next step prior to deposition, was to plasma clean substrates by inverse sputter etching with argon to ensure all surface contaminants are removed and finally transported *in vacuo* to the deposition chamber.

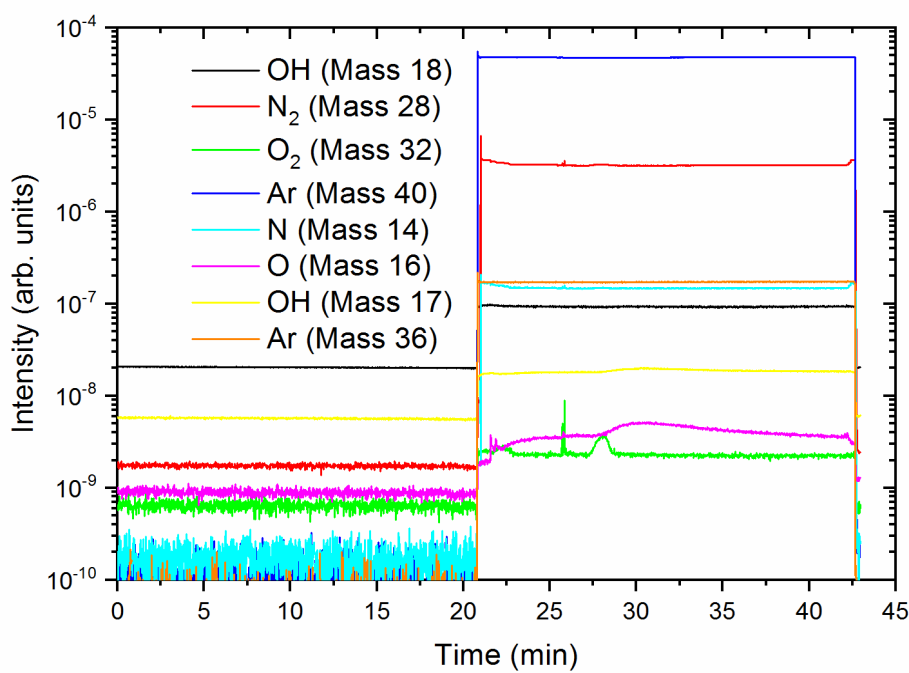

*Figure S1.1 : Residual Gas Analysis data for TiN deposition*
Residual Gas Analysis (RGA) data collected during each deposition indicates that the $O_2$ and OH partial pressures within each run were kept below $3\times10^{-5}$ Pa and $7.5\times10^{-4}$ Pa respectively. The data for the TiN deposition process is shown in Figure S1.1. Process gases Ar and $N_2$ are admitted at 21 minutes resulting in increased content of oxygen and water vapour in the chamber. When plasma power is turned off at the end of the process there is a slight increase in $N_2$ content due to the end of the reaction with the metal target. This is followed shortly by a switch off of the gas flow and the stop of data collection.

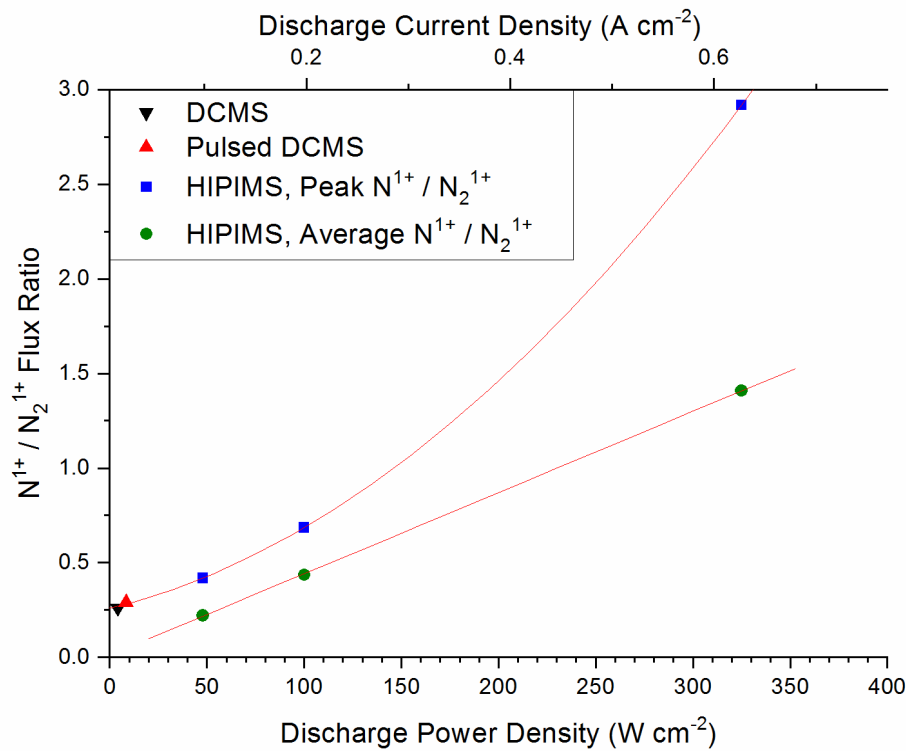

*Figure S1.2: Plasma sampling mass spectroscopy of TiN deposition. Squares indicate the ratio obtained at the peak of the HIPIMS pulse. Circles indicate the ratio obtained as a time-averaged measurement through the duration. Triangles indicate values for DC magnetron sputtering and pulsed DC magnetron sputtering.*

Plasma sampling mass spectroscopy of TiN deposition processes showed that the discharge power density resulted in an exponential increase in the $N^{1+} / N_2^{1+}$ ion flux ratio observed during the peak of the pulse (Figure S1.2). The same ratio collected as an average throughout the pulse increased linearly with power density. At powers below 200 Wcm$^{-2}$, corresponding to low HIPIMS powers as well as DC magnetron sputtering (~10 Wcm$^{-2}$) and pulsed DC magnetron sputtering (~20 Wcm$^{-2}$), the dissociation rate of nitrogen is insufficient. At high HIPIMS power, the dissociation rate is enhanced by an order of magnitude compared to conventional DC magnetron sputtering due to a comparable increase in plasma density and a higher probability of dissociative collisions.

2. <u>Stoichoimetric analysis</u>
    a. <u>Energy dispersive X-ray spectroscopy</u>

| Sample | Ti at.% | Nb at.% |
|---|---|---|
| TiNbN | 49.7 | 50.3 |
| TiN/NbN 10nm layers | 52.4 | 47.6 |
| TiN/NbN 5nm layers | 52.75 | 47.25 |

*Table S2.1: EDX stoichiometries for Ti and Nb in layered and co-sputtered films*

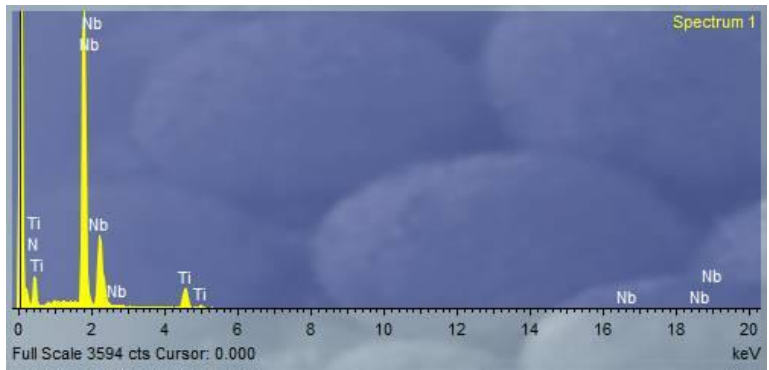

*Figure S2.1: EDX spectrum for TiNbN on Si*

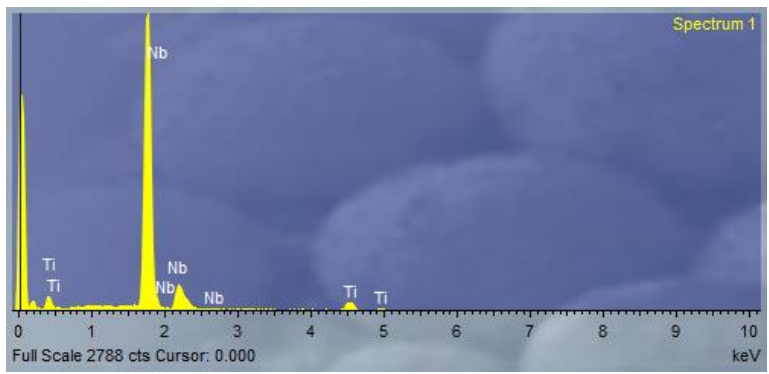

*Figure S2.2: EDX spectrm for TiN/NbN on Si (10nm layers)*

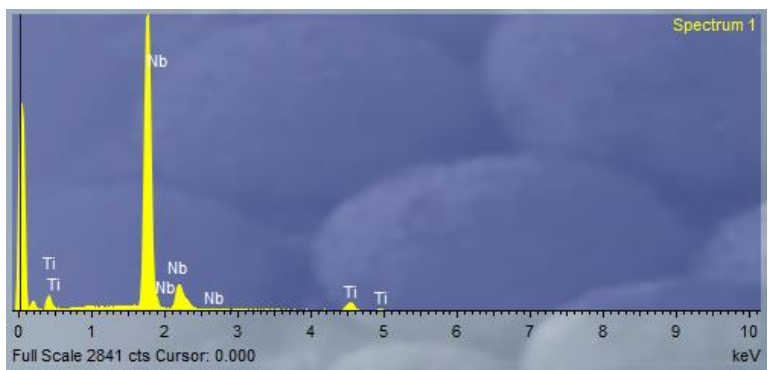

*Figure S2.3: EDX spectrum for TiN/NbN on Si (5nm layers)*

b. <u>XRF spectrum for co-sputtered thin films and binary thin films film</u>

| Sample | Ti at. % | Nb at. % |
|---|---|---|
| TiN | 99.1% | 0.1% |
| NbN | 2.9% | 97.1% |
| TiNbN | 48.1% | 51.9% |

*Table S2.2: XPS stoichiometries for Ti and Nb in binary and ternary films*

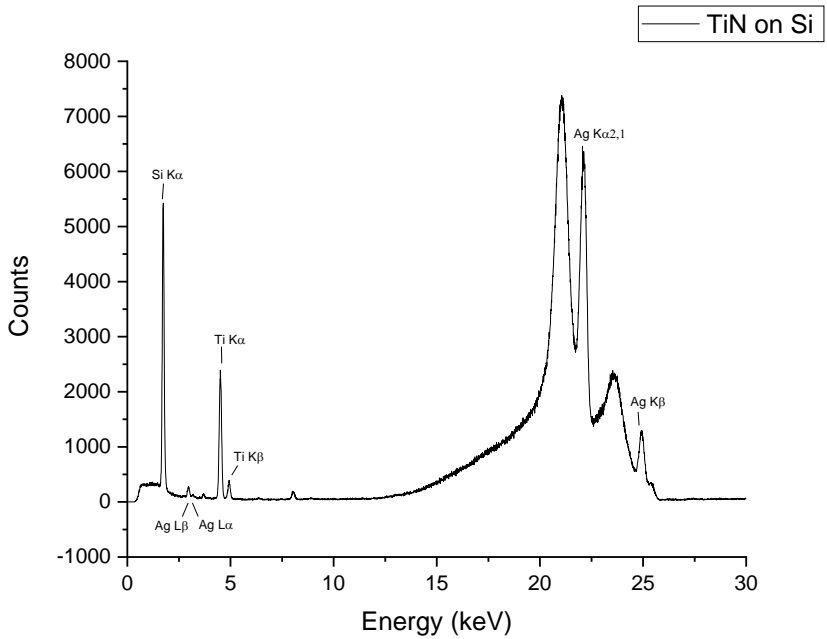

*Figure S2.4: XPS spectrum for TiN on Si*

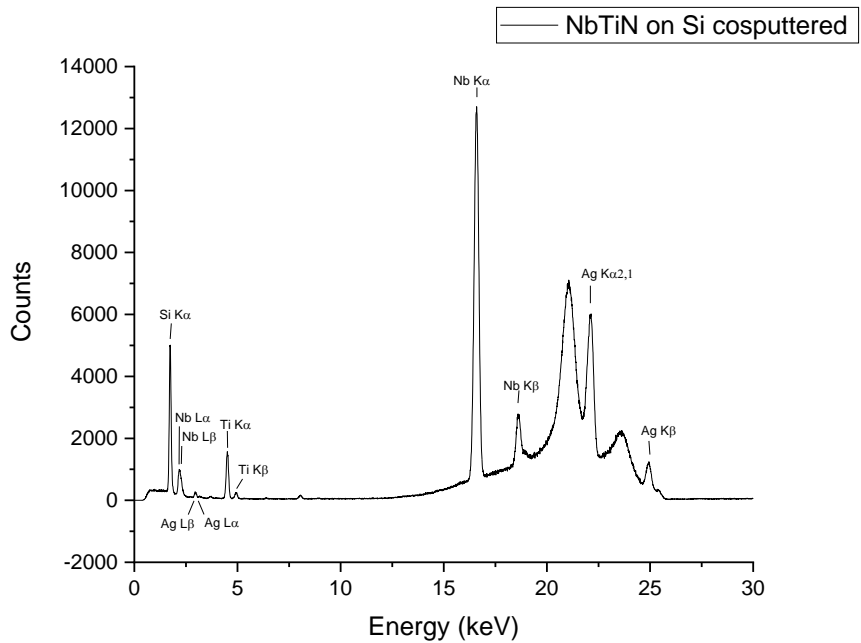

*Figure S2.5: XPS spectrum for TiNbN on Si*

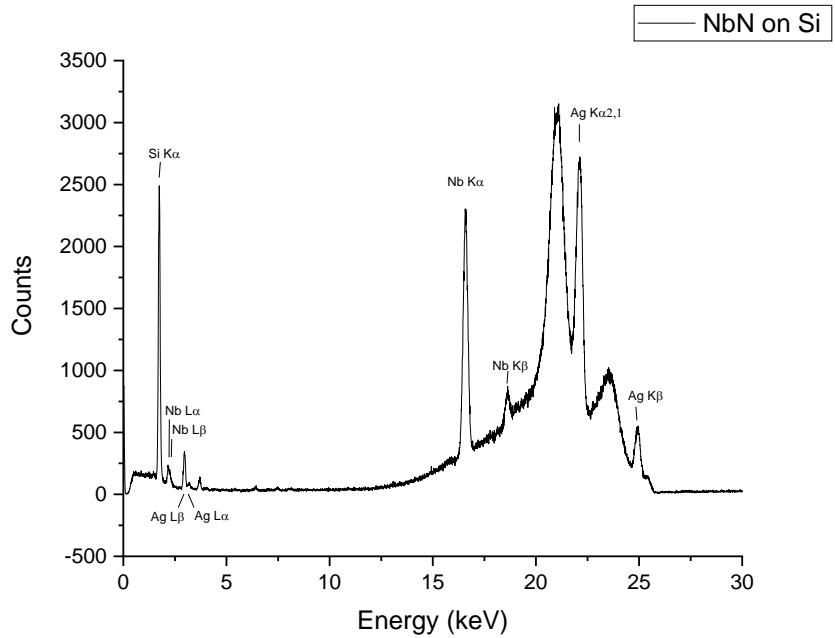

*Figure S2.6: XPS spectrum for NbN on Si*

The sharp Ag peaks visible in the XRF data are due to the primary Ag source used. Additionally, the broad peaks observed at approximately 20 keV and 24 keV are a result of Compton scattering. The presence of Ti in the NbN and Nb in the TiN films is an artefact from the measurement software and is indicative of the accuracy of the technique.

3. <u>Structural analysis (XRD)</u>

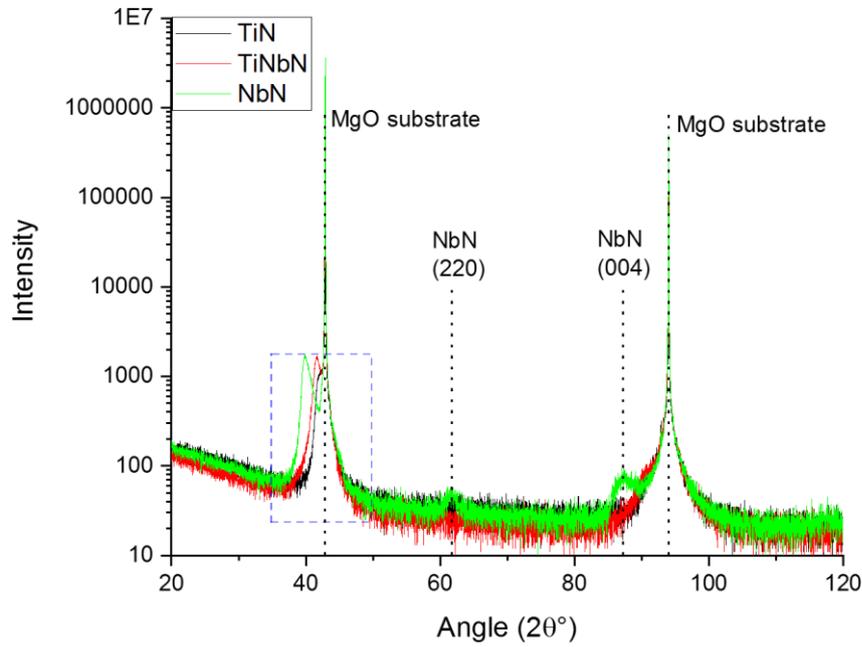

Figure S3.1: Thin film XRD of binary and ternary films on MgO

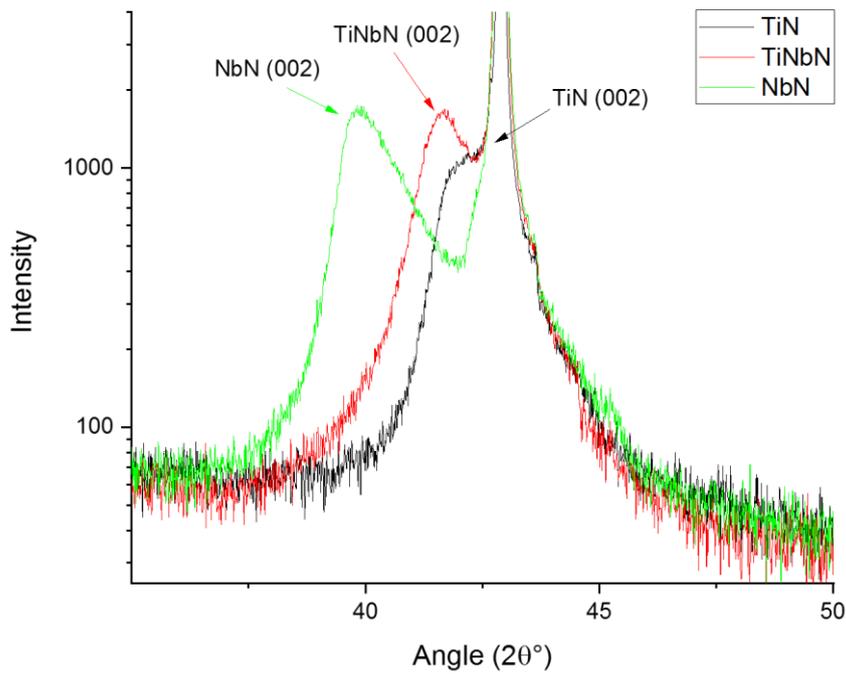

Figure S3.2: Thin film XRD of binary and ternary films (Inset)

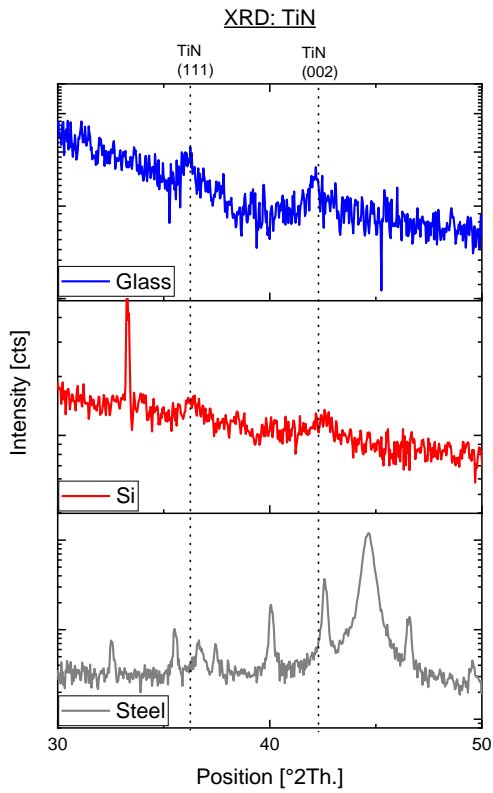

Figure S3.3: XRD data for TiN films films

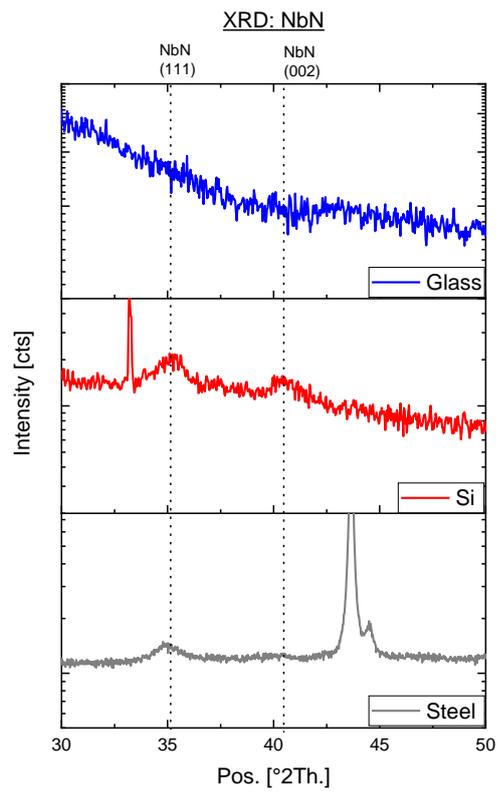

Figure S3.4: XRD data for NbN

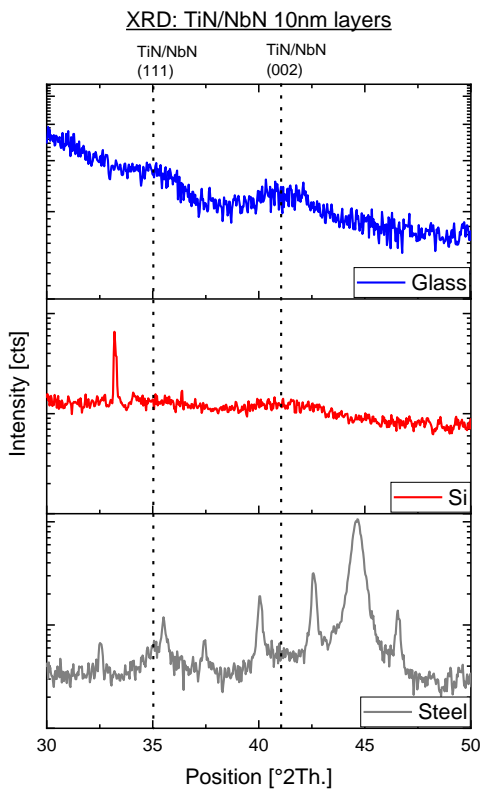

Figure S3.5: TiN/NbN layered films XRD data data

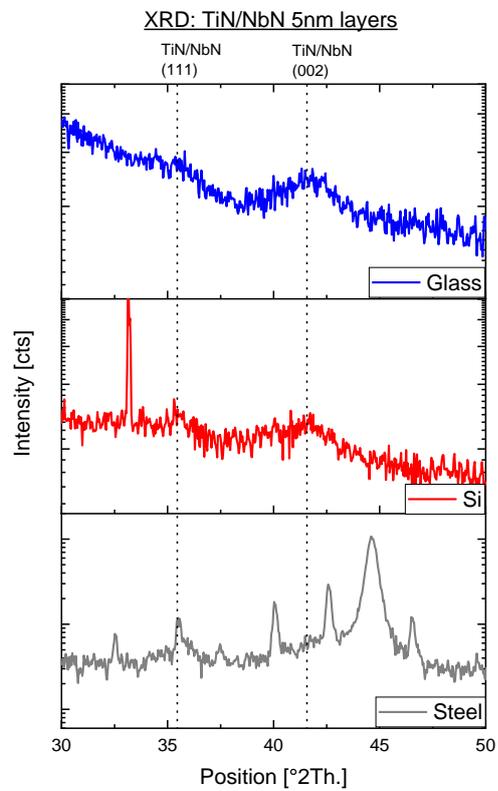

Figure S3.6: TiN/NbN layered films XRD data

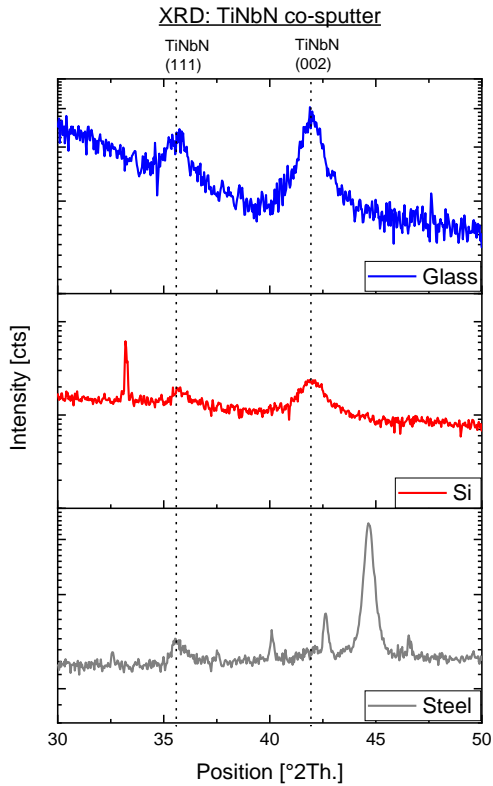

Figure S3.7: TiNNbN co-sputtered films XRD data

Table S3.1: Lattice parameters extracted from XRD data

| Sample | Substrate | Lattice Parameter [Å] |
|---|---|---|
| NbN | Glass | - |
|  | Si | 4.42 |
|  | Steel | 4.46 |
|  | MgO | 4.44 |
| TiN | Glass | 4.28 |
|  | Si | 4.29 |
|  | Steel | 4.25 |
|  | MgO | 4.24 |
| TiNbN | Glass | 4.32 |
|  | Si | 4.32 |
|  | Steel | 4.33 |
|  | MgO | 4.33 |
| TiN/NbN 5nm layers | Glass | 4.35 |
|  | Si | 4.35 |
|  | Steel | 4.39 |
| TiN/NbN 10nm layers | Glass | 4.39 |
|  | Si | - |
|  | Steel | 4.36 |

All peaks in the XRD data not indexed as belonging to the thin films are from the substrate.

4. <u>Surface Morphology: AFM</u>

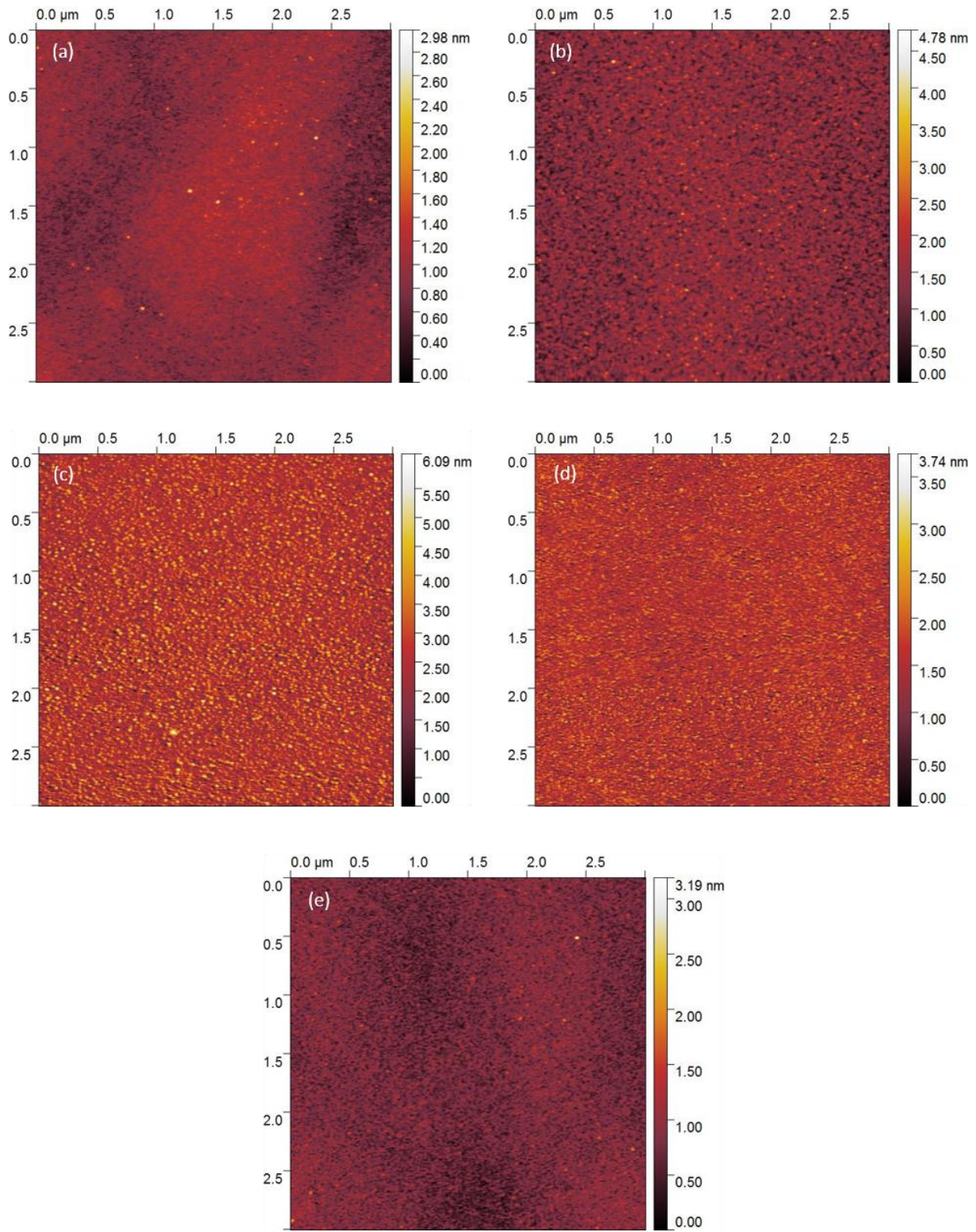

Figure S4: (a) NbN on Si. RMS = 0.189 nm; (b) TiNbN on Si. RMS = 0.35 nm; (c) TiN on Si. RMS = 0.666 nm; (d) TiN/NbN 10nm layers. RMS = 0.344 nm; (e) TiN/NbN 5nm layers. RMS = 0.195 nm

| Sample | RMS [nm] | Mean roughness [nm] |
|---|---|---|
| NbN | 0.189 | 0.150 |
| TiN/NbN layers (5nm) | 0.195 | 0.15 |
| TiN/NbN layers (10nm) | 0.344 | 0.270 |
| TiNbN | 0.35 | 0.28 |
| TiN | 0.666 | 0.52 |

Table S4.1: AFM surface roughness data for films on Si substrates

5. Ellipsometry Fitting

Ellipsometry data was fit using a Drude-Lorentz model of the form:

$$\varepsilon'(\omega) + i\varepsilon''(\omega) = \varepsilon_\infty - \frac{\omega_{pu}^2}{\omega^2 - i\Gamma_D} + \sum_{j=1}^{n} \frac{f_j \omega_{0_j}^2}{\omega_{0_j}^2 - \omega^2 + i\gamma_j \omega}$$

Where $\varepsilon_\infty$ relates to background absorption and $\omega_{pu}$ and $\Gamma_D$ are the unscreened plasma frequency and Drude broadening. The Lorentz oscillators are defined by the oscillator energy, strength, and damping: $\omega_j$, $f_j$, and $\gamma_j$ respectively.

A total of 2 Lorentz oscillators were used, as the inclusion of a third oscillator did not significantly improve the fit, as determined by the MSE.

Table S5.1: Ellipsometry fitting data for all films.

| Sample | Substrate | MSE | $\varepsilon_\infty$ | $E_{pu}$ [eV] | $\Gamma_D$ [eV] | $E_1$ [eV] | $f_1$ | $\gamma_1$ [eV] | $E_2$ [eV] | $f_2$ | $\gamma_2$ [eV] |
|---|---|---|---|---|---|---|---|---|---|---|---|
| TiN | Steel | 8.123 | 2.856 | 6.71 | 0.408 | 6.199 | 9.3 | 95 | 4.647 | 2.803 | 1.915 |
| | Glass | 6.461 | 3.506 | 7.26 | 0.338 | 5.644 | 12.26 | 96 | 4.669 | 2.806 | 1.813 |
| | Si | 5.557 | 3.558 | 7.44 | 0.416 | 4.802 | 3.819 | 2.4 | 2.024 | 0.997 | 1.617 |
| | MgO | 5.986 | 3.295 | 7.17 | 0.325 | 4.442 | 14.157 | 64.3 | 4.776 | 3.033 | 1.945 |
| NbN | Steel | 1.014 | 2.454 | 13.43 | 3.849 | 5.911 | 2.684 | 2 | 0.57 | 3.022 | 0.603 |
| | Glass | 3.019 | 2.430 | 12.67 | 3.849 | 8,662 | 5.216 | 6.6 | - | - | - |
| | Si | 1.441 | 1 | 10.58 | 3.222 | 5.666 | 2.291 | 2.03 | 3.035 | 7.174 | 5.5 |
| | MgO | 1.494 | 1 | 12.45 | 3.591 | 6.791 | 4.151 | 2.66 | - | - | - |
| TiNbN | Steel | 2.73 | 1 | 8.49 | 0.986 | 6.9 | 0.15 | 6.082 | - | - | - |
| | Glass | 1.613 | 2.319 | 8.71 | 0.920 | 9.78 | 2.08 | 45.1 | 5.739 | 4.051 | 3.37 |
| | Si | 1.516 | 1.896 | 8.67 | 0.954 | 6.158 | 5.547 | 4.525 | 2.254 | 0.224 | 1.71 |
| | MgO | 1.993 | 2.763 | 8.57 | 0.863 | 9.596 | 2.203 | 46.77 | 5.371 | 3.619 | 2.922 |
| TiN/NbN 10nm layers | Steel | 6.599 | 2.326 | 8.5 | 0.852 | 3.535 | 0.284 | 10 | 5.123 | 0.259 | 1.64 |
| | Glass | 6.792 | 2.251 | 9.06 | 0.744 | 3.016 | 0.567 | 10 | 5.482 | 0.244 | 1.951 |
| | Si | 4.22 | 2.624 | 8.71 | 0.826 | 3.091 | 0.471 | 10 | 5.339 | 0.23 | 2.092 |
| | MgO | 2.93 | 2.176 | 9.01 | 0.766 | 2.24 | 1.179 | 6.7 | 5.977 | 0.213 | 2.88 |
| TiN/NbN 5nm layers | Steel | 2.36 | 1.154 | 8.82 | 0.709 | 1.509 | 2.66 | 3.15 | 6.795 | 0.144 | 5.718 |
| | Glass | 1.98 | 1.046 | 9.08 | 0.623 | 1.277 | 7.343 | 3.26 | 7.466 | 0.133 | 6.336 |
| | Si | 1.847 | 2.193 | 9.16 | 0.629 | 1.302 | 6.713 | 1.3 | 6.262 | 0.173 | 4.699 |
| | MgO | 3.834 | 2.513 | 8.99 | 0.812 | 3.276 | 0.34 | 10 | 5.458 | 0.239 | 2.025 |

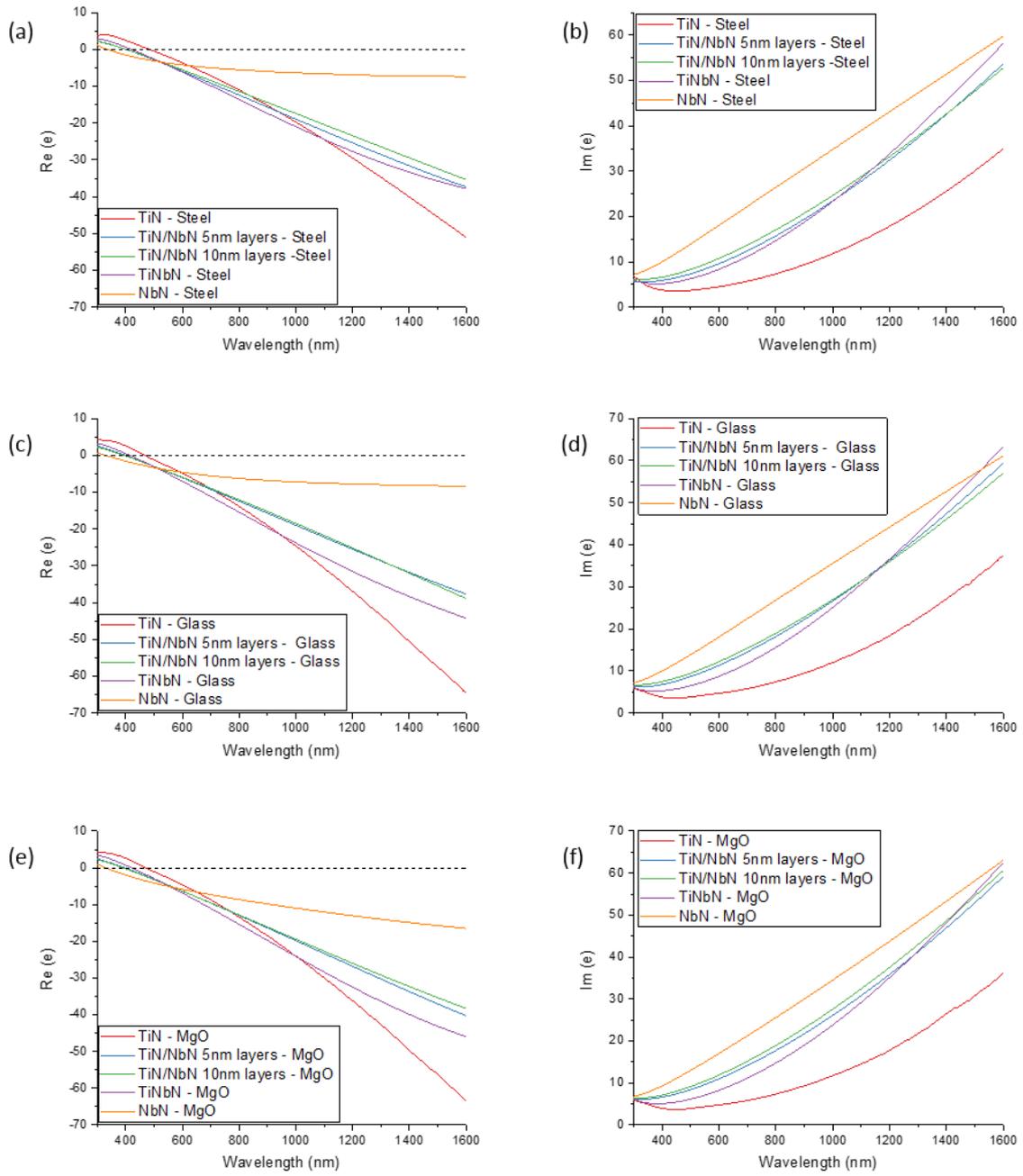

Figure S5.1: Real and imaginary permittivity for steel (a, b), glass (c, d) and MgO (e, f) substrate films